\documentclass[
    ,final            
  ]
  {aipproc}

\layoutstyle{8x11single}

\begin{document}

\title{Scalar, axial-vector and tensor resonances from the $\rho D^*$, $\omega
D^*$
interaction in the hidden gauge formalism}


\author{R. Molina}{
  address={Departamento de F\'{\i}sica Te\'orica and IFIC,
Centro Mixto Universidad de Valencia-CSIC,
Institutos de Investigaci\'on de Paterna, Aptdo. 22085, 46071 Valencia,
 Spain}
}

\author{H. Nagahiro}{ address={Department of Physics, Nara Women's University,
Nara 630-8506, Japan}
}

\author{A. Hosaka}{ address={Research Center for Nuclear Physics (RCNP), Osaka University,
Ibaraki, Osaka 567-0047, Japan}
}

\author{E. Oset}{
  address={Departamento de F\'{\i}sica Te\'orica and IFIC,
Centro Mixto Universidad de Valencia-CSIC,
Institutos de Investigaci\'on de Paterna, Aptdo. 22085, 46071 Valencia,
 Spain}
  } 

\begin{abstract}
 We apply a unitary approach together with a set of hidden-gauge Lagrangians to study the vector-vector interaction. We study the case of the $\rho$($\omega$)$D^*$ interaction within the model. In $I=1/2$ we get strong enough attraction to bind the system. Concretely, we get one resonance for each spin $J=0,1,2$. For $J=1$ and $2$ these resonances can be easily identified with the $D^*(2460)$ and $D^*(2640)$. Whereas the state for $J=0$ with mass $M\sim2600$ MeV is a prediction of the model. In $I=3/2$ we get a repulsive interaction and hence no exotic dynamically generated states.
\end{abstract}

\maketitle


\section{Introduction}
In this talk, we report several works that combine coupled channel unitarity and hidden-gauge Lagrangians for the interaction of vector meson among themselves. These Lagrangians were introduced by Bando-Kugo-Yamawaki \cite{hidden} and combined with unitary techniques related with the Bethe-Salpeter equation, provide a useful tool to generate resonances or bound states if the interaction is strong enough. These techniques were firstly applied to study the $\rho\rho$ interaction and the interaction was strong enough to bind the $\rho\rho$ system \cite{raquel}. Thus, two bound states appeared as poles in the second Riemann sheet that could be identified with the $f_2(1270)$ and the $f_0(1370)$. The decay of these resonances to two or four pions was provided by a box or crossed box diagram from two $\rho$ mesons going to pions. These box diagrams provided a width comparable with the data quoted at the PDG \cite{pdg}, leading to satisfactory results on the mass, width and quantum numbers, as well as couplings, and thus, these dynamically generated resonances can be seen as hadronic molecules. Later works that we quote in the next Section have extended the model to study the vector-vector interaction in SU(3) and the $\rho$($\omega$)$D^*$ system. 

In this manuscript we want to summarize briefly the formalism of the vector-vector interaction. Then, we want to show the simple case of the $\rho$($\omega$) $D^*$ interaction. 

\section{Formalism: The $VV$ interaction}
\begin{figure}[htb]
\centering
\includegraphics*[width=8cm]{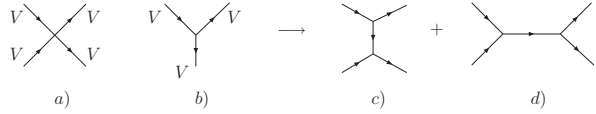}
\caption{Terms of the ${\cal L}_{III}$ Lagrangian: a) four vector contact term,
 Eq.~(\ref{lcont}); b) three-vector interaction, Eq.~(\ref{l3V}); c) $t$ and 
 $u$ channels from vector exchange; d) $s$ channel for vector exchange.}
\label{fig:3V4V}
\end{figure}

Within the theorical framework, there are two main ingredients: first, we take the Lagrangians for the interaction of vector mesons among themselves, that come from the hidden gauge formalism of Bando-Kugo-Yamawaki \cite{hidden}. Second, we introduce the potential $V$ obtained from these Lagrangians (projected in s-wave, spin and isospin) in the Bethe Salpeter equation:
\begin{equation}
T= (\hat{1}-VG)^{-1} V\ ,
\label{Bethe}
\end{equation}
where $G$ is the loop function. Therefore, we are summing all the diagrams containing zero, one, two... loops implicit in the Bethe Salpeter equation. Finally, we look for poles of the unitary $T$ matrix in the second Riemann sheet. All this procedure is well explained in \cite{raquel,geng}. In these works, two main vertices are taken into account for the computation of the potential $V$: the four-vector-contact term and the three-vector vertex, which are provided respectively from the Lagrangians:
\begin{equation}
{\cal L}^{(c)}_{III}=\frac{g^2}{2}\langle V_\mu V_\nu V^\mu V^\nu-V_\nu V_\mu
V^\mu V^\nu\rangle\ ,
\label{lcont}
\end{equation}
\begin{equation}
{\cal L}^{(3V)}_{III}=ig\langle (\partial_\mu V_\nu -\partial_\nu V_\mu) V^\mu V^\nu\rangle
\label{l3V}\ .
\end{equation}
By means of Eq. (\ref{l3V}), the vector exchange diagrams in Fig. \ref{fig:3V4V} are calculated. The diagram in Fig. \ref{fig:3V4V}d) leads to a repulsive p-wave interaction for equal masses of the vectors \cite{raquel} and only to a minor component of s-wave in the case of different masses \cite{geng}. Thus, the four-vector-contact term in Fig. \ref{fig:3V4V}a) and the t(u)-channel vector exchange diagrams in Fig. \ref{fig:3V4V}c) are responsible for the generation of resonances or bound states if the interaction is strong enough. 

\begin{figure}[htb]
\centering
\includegraphics*[width=8cm]{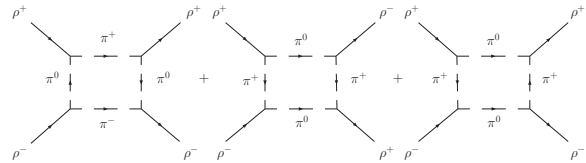}
\caption{Box diagrams for the case of the $\rho\rho$ interaction.}
\label{fig:box}
\end{figure}

In order to consider the pseudoscalar-pseudoscalar decay mode, and thus compare properly with the available experimental data for the width, a box diagram is also included, which we show in Fig. \ref{fig:box} for the particular case of the $\rho\rho$ system. Crossed box diagrams and box diagrams involving anomalous couplings were also calculated in \cite{raquel}, but they were much smaller, specially in the case of the anomalous coupling, than the contributions coming from the box diagrams in Fig. \ref{fig:box} and they were not considered in later works \cite{geng,raquel2,raquel3}. As we are interested in the region close to the two vector meson threshold, the three momenta of the external particles can be neglected compared with the mass of the vector meson, and thus, we can make $|\vec{q}|/M\sim 0$ for external particles, which considerably simplify the calculation. In Table \ref{tab:geng}, the eleven states found with this procedure in the work of \cite{geng}, where the formalism is applied in SU(3), are shown. As one can see in this table, five of them can be identified with data in the PDG: the $f_0(1370)$, $f_0(1710)$, $f_2(1270)$, $f_2'(1525)$ and $K^*_2(1430)$. In this work, the integral of two-meson-vector loop function is calculated by means of the dimensional regularization method. This requires the introduction of one parameter, which is the subtraction constant, that is tuned to reproduce the mass of the tensor states, whereas the other states are predicted. 
In order to calculate the box diagrams, one has to introduce two parameters, the cut-off, $\Lambda$, for the integral, and one parameter, $\Lambda_b$, involved in the form factor used. These parameters were considered to be around $1$ GeV and $1.4$ GeV respectively in \cite{raquel} to get reasonable values of the $f_0(1370)$ and $f_2(1270)$ widths. In the later work of \cite{geng}, these values of $\Lambda$ and $\Lambda_b$ also provide a good description of the widths for the other states, as shown in Table \ref{tab:geng}.

As one can see, the model in SU(3) provides a good description of the properties of many physical states. In the next sections, we generalize the model to SU(4), where interesting new states will appear.

\begin{table}
\begin{tabular}{lrrrr}
\hline
\tablehead{1}{l}{b}{$I^{G}(J^{PC})$}
  & \tablehead{1}{r}{b}{Theory}
  & \tablehead{1}{r}{b}{}
  & \tablehead{1}{r}{b}{PDG data}
  & \tablehead{1}{r}{b}{}   \\
\hline
              & (Mass, Width) & Name & Mass & Width  \\
$0^+(0^{++})$ & ($1520$, $257-396$) & $f_0(1370)$ & $1200\sim1500$ & $200\sim500$\\
$0^+(0^{++})$ & ($1720$, $133-151$) &$f_0(1710)$ & $1724\pm7$ & $137\pm 8$\\
$0^-(1^{+-})$ & ($1802$, $49$)  & $h_1$& & \\
$0^+(2^{++})$ & ($1275$, $97-111$) & $f_2(1270)$ & $1275.1\pm1.2$ & $185.0^{+2.9}_{-2.4}$\\
$0^+(2^{++})$ & ($1525$, $45-51$)  &$f_2'(1525)$ & $1525\pm5$ & $73^{+6}_{-5}$\\\hline
$1^-(0^{++})$   & ($1777$, $148-172$)  & $a_0$& & \\
$1^+(1^{+-})$    & ($1703$, $188$)  & $b_1$& & \\
$1^-(2^{++})$  &  ($1567$, $47-51$)& $a_2(1700)??$& &
\\\hline
$1/2(0^+)$      &  ($1639$, $139-162$) &  $K_0^*$& & \\
$1/2(1^+)$      & ($1743$, $126$) &   $K_1(1650)?$& &\\
$1/2(2^+)$       &  ($1431$, $56-63$)  &$K_2^*(1430)$ & $1429\pm 1.4$ & $104\pm4$\\
 \hline
    \end{tabular}
\caption{Properties of the 11 dynamically generated states: pole positions, masses and widths in the real axes for two different $\Lambda_b$ (parameter related to the form factor used in the calculation of the box diagram \cite{geng}) compared with the experiment. All the quantities are in units of MeV.}
\label{tab:geng}
\end{table}

\section{ The $\rho$($\omega$) $D^*$ system}
In this section we describe the simple case of the channels with $C=1$ and $S=0$ within the formalism of the vector-vector interaction introduced by Bando-Kugo-Yamawaki \cite{hidden}. We have a set of three channels: $\rho D^*$, $\omega D^*$ and $D_s\bar{K}^*$. In fact, we want to focus the attention to the energy region around $2500-2600$ MeV, where there are several interesting resonances that can be found in the PDG. In view of the previous works of \cite{raquel,geng}, we hope to find a binding energy of $\sim200-300$ MeV and the threshold of the $D_s\bar{K}^*$ channel is far away, concretely, it is $200$ MeV above the $\rho D^*$ and $\omega D^*$ thresholds. Thus, we simplify even more the study to these two channels: $\rho D^*$ and $\omega D^*$. 

In order to follow the procedure of the works of \cite{raquel,geng}, the $V$ matrix is straigthforward extended to SU(4), as it was done in the work of \cite{Gamermann}:
\begin{equation}
V_\mu=\left(
\begin{array}{cccc}
\frac{\rho^0}{\sqrt{2}}+\frac{\omega}{\sqrt{2}}&\rho^+& K^{*+}&\bar{D}^{*0}\\
\rho^-& -\frac{\rho^0}{\sqrt{2}}+\frac{\omega}{\sqrt{2}}&K^{*0}&D^{*-}\\
K^{*-}& \bar{K}^{*0}&\phi&D^{*-}_s\\
D^{*0}&D^{*+}&D^{*+}_s&J/\psi\\
\end{array}
\right)_\mu \ ,
\label{Vmu}
\end{equation}
where the ideal mixing has been taken for $\omega$, $\phi$ and $J/\psi$. The details of the full procedure that we try to summarize in this section are well explained in \cite{raquel2}. Besides the four-vector contact diagrams, we have two kinds of vector-exchange diagrams, one is the exchange of one light vector meson, $\rho$ or $\omega$, and the other one is the exchange of a heavy vector meson, $D^*$. Of course, the last terms are proportional to $\kappa=\frac{m^2_\rho}{m^2_{D^*}}\sim 0.15$, and this gives rise to corrections of the order of $10$\% of the $\rho$-exchange terms. Also, the $\rho\rho \omega$ and the $\omega\omega\omega$ vertices violate G-parity, whereas the $\rho\omega\omega$ vertex violates isospin, therefore, we do not have the exchange of one light vector meson in the $V(\rho D^*\to \omega D^*)$ and $V(\omega D^*\to \omega D^*)$ potentials. For this reason, these two terms, where only the four-vector contact term plus the $D^*$-exchange term contributes , are smaller compared with the $V(\rho D^*\to \rho D^*)$, term that mainly provides the interaction. As can be seen in Table \ref{tab:pot}, this term is of the order of $-300$ MeV in $I=1/2$. This magnitude is responsible to bind the $\rho D^*$ system and comes mainly from the $\rho$-exchange term of $V(\rho D^*\to \rho D^*)$. This minor relevance of the $\omega D^*$ channel appears when one calculate the couplings $g_{i}$ to one particular channel $i$ by means of the residues of the amplitudes \cite{raquel2} that we show in Table \ref{tab:coup}. In $I=3/2$, it is interesting to see that we get a repulsive interaction without any possible generated exotic states. 

\begin{table}
\begin{tabular}{lrrrrr}
\hline
    \tablehead{1}{l}{b}{$I$}
  & \tablehead{1}{r}{b}{$J$}
  & \tablehead{1}{r}{b}{Contact}
  & \tablehead{1}{r}{b}{$\rho$-exchange}
  & \tablehead{1}{r}{b}{$D^*$-exchange}
  & \tablehead{1}{r}{b}{$\sim$ Total$[I(J^{P})]$}   \\
\hline
$1/2$&$0$&$+5g^2$&$-2\frac{g^2}{M_{\rho}^2}\,(k_1+k_3)\cdot (k_2 +k_4)$&$-\frac{1}{2}\frac{\kappa\, g^2}{M_{\rho}^2}\,(k_1+k_4)\cdot (k_2 +k_3) $&$-16g^2[1/2(0^{+})]$\\
$1/2$&$1$&$+\frac{9}{2}g^2$&$-2\frac{g^2}{M_{\rho}^2}\,(k_1+k_3)\cdot (k_2 +k_4)$&$+\frac{1}{2}\frac{\kappa\, g^2}{M_{\rho}^2}\,(k_1+k_4)\cdot (k_2 +k_3) $&$-14.5g^2[1/2(1^{+})]$\\
$1/2$&$2$&$-\frac{5}{2}g^2$&$-2\frac{g^2}{M_{\rho}^2}\,(k_1+k_3)\cdot (k_2 +k_4)$&$-\frac{1}{2}\frac{\kappa\, g^2}{M_{\rho}^2}\,(k_1+k_4)\cdot (k_2 +k_3) $&$-23.5g^2[1/2(2^{+})]$\\
$3/2$&$0$&$-4g^2$&$+\frac{g^2}{M_{\rho}^2}\,(k_1+k_3)\cdot (k_2 +k_4)$&$+\frac{\kappa\, g^2}{M_{\rho}^2}\,(k_1+k_4)\cdot (k_2 +k_3) $&$+8g^2[3/2(0^{+})]$\\
$3/2$&$1$&$0$&$+\frac{g^2}{M_{\rho}^2}\,(k_1+k_3)\cdot (k_2 +k_4)$&$-\frac{\kappa\, g^2}{M_{\rho}^2}\,(k_1+k_4)\cdot (k_2 +k_3) $&$+8g^2[3/2(1^{+})]$\\
$3/2$&$2$&$+2g^2$&$+\frac{g^2}{M_{\rho}^2}\,(k_1+k_3)\cdot (k_2 +k_4)$&$+\frac{\kappa\, g^2}{M_{\rho}^2}\,(k_1+k_4)\cdot (k_2 +k_3) $&$+14g^2[3/2(2^{+})]$\\
\hline
\end{tabular} 
\caption{$V(\rho D^*\to \rho D^*)$ for the different spin-isospin channels including the exchange of one 
heavy vector meson. The approximate Total is obtained at the threshold of $\rho D^*$.}
\label{tab:pot}
\end{table} 
\begin{table}
\begin{tabular}{lrrr}
\hline
   \tablehead{1}{l}{b}{Channel}
  & \tablehead{1}{r}{b}{$D_0^*(2600)$}
  & \tablehead{1}{r}{b}{$D_1^*(2640)$}
  & \tablehead{1}{r}{b}{$D_2^*(2460)$}
\\
\hline
$\rho D^*$&14.32&14.04&17.89\\
$\omega D^*$&0.53&1.40&2.35\\
\hline
\end{tabular}
\caption{Modules of the couplings $g_i$ in units of GeV for the poles in
 the $J=0,\,1,\,2;\ I=1/2$ sector with the channel $\rho D^*$ and $\omega D^*$.} 
 \label{tab:coup}
\end{table}
We get three states with $I=1/2$ and $J=0,1$ and $2$ respectively. We have fixed the value of $\mu$ as $1500$ MeV and we have fine-tuned the subtraction constant $\alpha$ around its natural value of $-2$ \cite{oller,gamerman2}, in order to get the position of the $D^*_2(2460)$ state at the PDG. Thus, we have chosen the value of $\alpha=-1.74$. Because of the strong interaction we get three bound states with practically no width except by the small width provided by the convolution of the $\rho$ propagator to take into account its decay in two pions. But, the later inclusion of the $\pi D$ box diagram allows us to get some more width comparable with the data in the PDG as can be seen in Table \ref{tab:thexp1}. In order to calculate the $\pi D$-box diagram, two different form factors for the vector-two-pseudoscalar vertex are considered. One is used in the works of \cite{raquel,geng}, which is inspired
by the empirical form factors used in the study of vector meson decays \cite{titov}. The other one, is an exponential parametrization for an off-shell pion evaluated using QCD sum rules \cite{navarra} using at the same time the experimental value of the $D^*D\pi$ coupling, $g^{exp}_{D^*D\pi}$ \cite{raquel2}. In both cases we get reasonable values of the width, with the preference, of course, for the use of $g^{exp}_{D^*D\pi}$. In Table \ref{tab:thexp1}, the values obtained for the last case are shown, where the parameter of the exponential form factor is taken as $\Lambda=1$ GeV. The interesting thing in the calculation of the box diagram, is that this box diagram only has $J=0$ and $2$, therefore, the state found with $J=1$ cannot decay to $\pi D$ by means of this mechanism, which is the reason why we get a small width  of $4$ MeV, compared with the other states for $J=0$ and $2$. Thus, we find a reasonable 
explanation for why the $D^*(2640)$ has a small width, $\Gamma<15$ MeV, when we associate to this state the quantum numbers $J^{P}=1^{+}$. 

In Table \ref{tab:uncer} we show the uncertainties related to the SU(4) breaking due to the use of the parameter $g=m_V/2\,f$ in the Lagrangian. In this table we distinguish three different cases in which we use $g^2$, $g\,g_D$ or $g_D^2$ in the amplitudes obtained, see Table \ref{tab:pot}, with $g=m_{\rho}/2\,f_\pi$ and $g_D=m_{D^*}/2\,f_D$, and fixing the $\alpha$ parameter to get the pole position of the tensor state. This is always done and therefore this method provides a realistic way of looking at the uncertainties. We can see in this table that the changes in the mass of the other states are around $20$ MeV, which is quite small, and even if one fixes alpha and looks for the uncertainties using $g^2$, $g\,g_D$ or $g_D^2$, then the changes in the mass are around $70-90$ MeV, which are typical in any hadron model.
\begin{table}
\begin{tabular}{lrrrr}
\hline
    \tablehead{1}{l}{b}{$I^G[J^P]$}
  & \tablehead{1}{r}{b}{Theory}
  & \tablehead{1}{r}{b}{}
  & \tablehead{1}{r}{b}{PDG data}
  & \tablehead{1}{r}{b}{}  
\\\hline
              & (Mass, Width) & Name & Mass & Width  \\
$1/2^+(0^{+})$ & ($2608$, $61$) & "$D^*_0(2600)$" &  & \\
$1/2^+(1^+)$ & ($2620$, $4$) &$D^*(2640)$ & $2637$ & $<15$\\
$1/2^+(2^+)$ & ($2465$, $40$)  & $D^*(2460)$&$2460$ &$37-43$ \\
 \hline
    \end{tabular}
\caption{Masses and widths (in units of MeV) obtained in the case of the use of an exponential form factor with $\Lambda=1$ GeV compared with the experiment.}
\label{tab:thexp1}
\end{table}
\begin{table}
\begin{tabular}{lrrr}
\hline
    \tablehead{1}{l}{b}{Constant \& $\alpha$}
  & \tablehead{1}{r}{b}{$J=0$}
  & \tablehead{1}{r}{b}{$J=1$}
  & \tablehead{1}{r}{b}{$J=2$}
 \\
\hline
$g^2$ \& $-1.74$&$2592$&$2611$&$2450$\\
$g g_D$ \& $-1.53$&$2571$&$2587$&$2450$\\
$g_D^2$ \& $-1.39$&$2551$&$2565$&$2450$\\
\hline
\end{tabular}
\caption{Pole positions and subtraction constant obtained for the three different cases, $g^2$, $g g_D$ and $g_D^2$, when one fixes the mass of the pole with $J=2$.}
\label{tab:uncer}
\end{table} 

\section{Conclusions}
We have studied the $\rho(\omega)D^*$ interaction combining coupled channel unitarity and the hidden gauge Lagrangians for the interaction of vector mesons. We find a strong interaction that provides one bound state for each spin $J=0,1$ and $2$. We can assign the states with $J=1$ and $2$ to the $D^*(2640)$ and $D^*_2(2460)$ respectively, whereas the state obtained for $J=0$ is a prediction of the model: '$D_0^*(2600)$'. In this scheme, the $D^*_2(2460)$ and the $D^*_0(2600)$ decay to $\pi D$ by means of a box diagram. However, this decay is not possible for $J=1$, providing a natural explanation on why, whereas this state is heavier than the $D^*_2(2460)$, its width found in the PDG is very small compared to that, $\Gamma<15$ MeV, and the decay to $\pi D$ has not been observed either. 
The results obtained here should stimulate the search for more $D$ states in the region of $2600\,MeV$.  

\begin{theacknowledgments}

This work is partly supported by DGICYT contract number
FIS2006-03438. We acknowledge the support of the European Community-Research Infrastructure
Integrating Activity
Study of Strongly Interacting Matter (acronym HadronPhysics2, Grant Agreement
n. 227431)
under the Seventh Framework Programme of EU.
A.~H. is supported in part by the Grant for Scientific Research
Contract No.~19540297 from the  Ministry of Education, Culture,
Science and Technology, Japan.  
H.~N. is supported by the Grant for Scientific Research
No.~18-8661 from JSPS.

\end{theacknowledgments}
\bibliographystyle{aipproc}

\end{document}